\documentclass[prd,superscriptaddress,preprintnumbers,twocolumn,nofootinbib,showpacs]{revtex4-2}

\usepackage{amsfonts,amssymb,amsmath}
\usepackage{mathrsfs}
\usepackage{color}
\usepackage{graphicx}
\usepackage{dcolumn}
\usepackage{bm}

\newcommand{\half}{\frac 12}

\newcommand{\Slash}[1]{{\ooalign{\hfil#1\hfil\crcr\raise.167ex\hbox{/}}}}

\begin{document}

\title{Primordial blackholes from Gauss-Bonnet-corrected single field inflation}
\author{Shinsuke Kawai}
\email{kawai@skku.edu}
\affiliation{
	Department of Physics, 
	Sungkyunkwan University,
	Suwon 16419, Republic of Korea}
\author{Jinsu Kim}
\email{jinsu.kim@cern.ch}
\affiliation{
	Theoretical Physics Department,
	CERN,
	1211 Geneva 23, Switzerland} 
\date{\today} 
\preprint{CERN-TH-2021-115}

\begin{abstract}
Primordial blackholes formed in the early Universe via gravitational collapse of over-dense regions may contribute a significant amount to the present dark matter relic density. 
Inflation provides a natural framework for the production mechanism of primordial blackholes. 
For example, single field inflation models with a fine-tuned scalar potential may exhibit a period of ultra-slow roll, during which the curvature perturbation may be enhanced to become seeds of the primordial blackholes formed as the corresponding scales reenter the horizon. 
In this work, we propose an alternative mechanism for the primordial blackhole formation. 
We consider a model in which a scalar field is coupled to the Gauss-Bonnet term and show that primordial blackholes may be seeded when a scalar potential term and the Gauss-Bonnet coupling term are nearly balanced.
Large curvature perturbation in this model not only leads to the production of primordial blackholes but it also sources gravitational waves at the second order. 
We calculate the present density parameter of the gravitational waves and discuss the detectability of the signals by comparing them with sensitivity bounds of future gravitational wave experiments.
\end{abstract}

\maketitle

\section{Introduction}

Primordial blackholes \cite{Zeldovich:1967lct,Hawking:1971ei,Carr:1974nx,Polnarev:1985btg} are a viable candidate for dark matter \cite{Chapline:1975ojl,Ivanov:1994pa} and may constitute all or a part of the dark matter relic density today. 
Recent studies of primordial blackholes as a dark matter candidate include Refs. \cite{Blais:2002nd,Afshordi:2003zb,Frampton:2010sw,Belotsky:2014kca,Carr:2016drx,Inomata:2017okj,Belotsky:2018wph,Carr:2020xqk}.
Unlike astrophysical blackholes, primordial blackholes may form in the early Universe through the gravitational collapse of large over-dense regions; see, e.g., Refs. \cite{Khlopov:2008qy,Sasaki:2018dmp,Carr:2020gox,Carr:2020xqk,Green:2020jor,Villanueva-Domingo:2021spv} for reviews.

Cosmic inflation provides a natural framework for the production of primordial blackholes. 
Single field inflation is capable of generating large primordial curvature perturbation in small scales compared to the scale of the cosmic microwave background \cite{Bullock:1996at,Yokoyama:1998pt,Leach:2000ea,Kohri:2007qn,Saito:2008em,Bugaev:2008gw,Alabidi:2009bk,Drees:2011hb,Drees:2011yz,Garcia-Bellido:2017mdw,Ezquiaga:2017fvi,Germani:2017bcs,Ballesteros:2017fsr}. 
Once the mode with large curvature perturbation reenters the horizon, gravitational collapse may occur, thereby forming primordial blackholes. 
Single field models with a canonical kinetic term and minimal coupling to gravity, however, require a severe fine-tuning of the scalar potential to generate an inflection point near the end of inflation that produces large enough density perturbations \cite{Hertzberg:2017dkh}. 
It has also been pointed out that the standard slow-roll approximation breaks down near the inflection point \cite{Hertzberg:2017dkh,Motohashi:2017kbs,Drees:2019xpp}.
In multifield inflation, e.g., in hybrid inflation models, generation of primordial blackholes is less challenging; see Refs. \cite{Silk:1986vc,Yokoyama:1995ex,Randall:1995dj,Garcia-Bellido:1996mdl,Kawasaki:1997ju,Kawasaki:2012wr,Kohri:2012yw,Clesse:2015wea,Ando:2017veq,Ketov:2019mfc,Braglia:2020eai,Ashoorioon:2020hln,Braglia:2020taf,Palma:2020ejf}.
Other models for generating large primordial curvature perturbation during inflation include the one with the modified gravity sector, and with a non-canonical form of the inflaton; see, e.g., Refs. \cite{Kannike:2017bxn,Pi:2017gih,Cheong:2019vzl,Lin:2020goi,Yi:2020kmq,Yi:2020cut,Gao:2020tsa,Chen:2021nio,Teimoori:2021thk,Heydari:2021gea}. In Ref. \cite{Ashoorioon:2019xqc} the production of primordial blackholes is discussed in the effective field theory framework.
It is also possible to produce primordial blackholes during (p)reheating \cite{Green:2000he,Bassett:2000ha,Martin:2019nuw,Auclair:2020csm}.

The enhancement in the curvature power spectrum not only seeds the primordial blackholes but it may also act as a source for the tensor perturbations at the nonlinear order, producing scalar-induced second-order gravitational waves \cite{Matarrese:1997ay,Mollerach:2003nq,Ananda:2006af,Baumann:2007zm}. 
Therefore, inflation models in which primordial blackholes are formed via large enhancement of the curvature perturbation inevitably produce the scalar-induced second-order gravitational waves. 
Detection or nondetection of gravitational wave signals may thus confirm or constrain the mass ranges of the primordial blackholes. 
For example, from the pulsar timing array experiments, a relatively large mass range of $[0.1M_{\odot},10M_{\odot}]$ ($M_{\odot}$ is the solar mass) is strongly constrained \cite{Lentati:2015qwp,Shannon:2015ect}. 
Recently, the NANOGrav Collaboration reported a hint for stochastic gravitational wave signals \cite{NANOGrav:2020bcs}. 
In Refs. \cite{Vaskonen:2020lbd,DeLuca:2020agl,Kohri:2020qqd,Sugiyama:2020roc,Domenech:2020ers,Inomata:2020xad}, the connection between the NANOGrav results and the primordial blackholes is investigated.
The future gravitational wave experiments such as Laser Interferometer Space Antenna (LISA) \cite{LISA:2017pwj,Baker:2019nia}, DECi-hertz Interferometer Gravitational wave Observatory (DECIGO) \cite{Seto:2001qf,Kawamura:2006up}, and Big Bang Observer (BBO) \cite{Crowder:2005nr,Corbin:2005ny,Harry:2006fi}, may probe smaller mass ranges of primordial blackholes.
Other recent studies on the primordial blackholes and the scalar-induced second-order gravitational waves include Refs. \cite{Sasaki:2016jop,Garcia-Bellido:2016dkw,Inomata:2016rbd,Domcke:2017fix,Garcia-Bellido:2017qal,Sasaki:2018dmp,Bhattacharya:2019bvk,Fu:2019vqc,Almeida:2020kaq,Ozsoy:2020kat,Papanikolaou:2020qtd,Domenech:2020ssp,Yuan:2021qgz,Franciolini:2021tla,Chen:2021nio,Teimoori:2021thk,Heydari:2021gea}.

In the single field inflation models for which the primordial blackhole production and the secondary gravitational waves are studied, the gravity sector is usually assumed to be the Einstein gravity. 
The Einstein gravity however is by no means a complete theory. 
From the effective field theory viewpoint, for example, higher curvature terms are expected to arise \cite{Weinberg:2008hq}. 
One such higher curvature term is the Gauss-Bonnet term $R_{\rm GB}^2 \equiv R^2 - 4R_{\mu\nu}R^{\mu\nu}+R_{\mu\nu\rho\sigma}R^{\mu\nu\rho\sigma}$ which leads to a relatively well-behaved theory of higher curvature gravity.
Phenomenological aspects of the Gauss-Bonnet correction have been studied by many authors, including dark energy \cite{Nojiri:2005vv,Koivisto:2006xf,Nojiri:2007te,Amendola:2007ni,Granda:2014zea}, inflation \cite{Kawai:1998ab,Kawai:1999pw,Satoh:2007gn,Satoh:2008ck,Guo:2009uk,Guo:2010jr,Satoh:2010ep,Koh:2014bka,Yi:2018gse,Nojiri:2019dwl,Odintsov:2019clh,Odintsov:2020sqy,Pozdeeva:2020shl,Oikonomou:2020sij,Pozdeeva:2020apf,Oikonomou:2020oil,Odintsov:2020xji,Odintsov:2020mkz,Oikonomou:2020tct}, blackholes \cite{Antoniou:2017acq,Lee:2018zym}, and gravitational-wave leptogenesis \cite{Kawai:2017kqt}.
In Ref. \cite{Kawai:2021bye}, we investigated a model in which a scalar field $\varphi$ is coupled to the Gauss-Bonnet term and discussed the features of a de Sitter-like fixed point as an alternative to cosmic inflation;
in the presence of the Gauss-Bonnet coupling term, there may exist a nontrivial de Sitter-like fixed point where the scalar potential term is balanced with the higher curvature Gauss-Bonnet term. 
Near the nontrivial fixed point, the standard slow-roll approximation is invalid, and the ultra-slow-roll regime of inflation naturally arises. 
Furthermore, we pointed out that the primordial curvature power spectrum may become enhanced near the nontrivial de Sitter-like fixed point, which potentially leads to production of primordial blackholes.
In this paper, we investigate the production of primordial blackholes and the scalar-induced second-order gravitational waves in such a setup.

The rest of the paper is organized as follows. 
In Sec. \ref{sec:model}, we introduce our benchmark model. 
We adopt the natural inflation model for the scalar potential and consider a smeared step function for the Gauss-Bonnet coupling function.
We choose two benchmark parameter sets and discuss enhancements in the primordial curvature power spectrum. 
In Sec. \ref{sec:PBH}, we examine the production of primordial blackholes. 
For the chosen benchmark sets, the produced primordial blackholes are shown to constitute all or a part of the dark matter relic today, at different mass scales. 
The two chosen benchmark sets give rise to the scalar-induced second-order gravitational wave signals peaked at two distinct frequencies. 
The shape and the magnitude of the gravitational wave energy density are discussed in Sec. \ref{sec:SIGW}, together with the detectability of the signals in the near future. 
We conclude in Sec. \ref{sec:conc} with some comments.

\section{Benchmark model}
\label{sec:model}

We consider the action
\begin{align}\label{eqn:action}
	S=\int d^4 x\sqrt{-g}\Big\{\frac{M_{\rm P}^2}{2}R
	-\half\partial_\mu\varphi\partial^\mu\varphi-V(\varphi)
	-\frac{\xi(\varphi)}{16}R_{\rm GB}^2
	\Big\}
\end{align}
where $M_{\rm P}\equiv 1/\sqrt{8\pi G}=2.44\times 10^{18}$ GeV is the reduced Planck mass, and $R_{\rm GB}^2\equiv R^2-4R_{\mu\nu}R^{\mu\nu}+R^{\mu\nu\rho\sigma}R_{\mu\nu\rho\sigma}$ is the Gauss-Bonnet term (the 4-dimensional Euler density).
The coupling function is chosen to be\footnote{
This form of the coupling is motivated in microscopic physics as follows.
In calculable examples of type II and heterotic compactifications, the $R_{\rm GB}^2$ coupling to moduli typically occurs as 1-loop gravitational threshold corrections, which are determined by the spectrum of BPS states \cite{Antoniadis:1992sa,Harvey:1995fq}.
If one of the moduli $\varphi$ traverses a wall separating two domains with different spectra, the coupling function $\xi(\varphi)$ behaves as a step function across the wall. 
If the domain wall has a finite thickness, $\xi(\varphi)$ is modeled by a smeared step function \eqref{eqn:xi}. Adding a constant parameter to Eq.~\eqref{eqn:xi} does not alter the dynamics since that will be a topological term.
}
\begin{align}\label{eqn:xi}
\xi(\varphi)&=\xi_0\tanh\left[\xi_1(\varphi-\varphi_c)\right]
\,,
\end{align}
and for the scalar potential we assume that of the natural inflation model \cite{Freese:1990rb,Adams:1992bn,Savage:2006tr,Freese:2014nla} (see also Refs. \cite{Kim:2004rp,Dimopoulos:2005ac,Grimm:2007hs}),
\begin{align}\label{eqn:pot}
V(\varphi) = \Lambda^4\left(1+\cos\frac{\varphi}{f}\right)\,.
\end{align}
The generation of primordial blackholes and the induced second-order gravitational waves in the natural inflation model \eqref{eqn:pot} (without the Gauss-Bonnet correction) has been discussed, e.g., in Refs. \cite{Bugaev:2013fya,Garcia-Bellido:2016dkw,Almeida:2020kaq,Domcke:2017fix,Ozsoy:2020kat}, where the axion (inflaton) field is coupled to gauge fields. 
References \cite{Gao:2020tsa,Teimoori:2021thk} discuss primordial blackholes and the gravitational waves in the natural inflation model with the modified inflaton kinetic sector.

The background equations of motion for the action \eqref{eqn:action} read
\begin{align}
&3M_{\rm P}^2H^2 = \frac{1}{2}\dot{\varphi}^2 + V + \frac{3}{2}H^3\xi_{,\varphi}\dot{\varphi}
\,,\label{eqn:bkgeom1}\\
&\ddot{\varphi} + 3H\dot{\varphi} + V_{,\varphi} + \frac{3}{2}H^2\left(\dot{H}+H^2\right)\xi_{,\varphi} = 0
\,,\label{eqn:bkgeom2}
\end{align}
where $_{,\varphi} \equiv d/d\varphi$ and $\dot{} \equiv d/dt$ with $t$ being the cosmic time.
$H \equiv \dot a/a$ is the Hubble parameter ($a$ is the scale factor).
The model \eqref{eqn:action} exhibits a nontrivial fixed point $\varphi_*$ that satisfies \cite{Kawai:2021bye}
\begin{align}
\left[
V_{,\varphi} + \frac{V^2}{6M_{\rm P}^4}\xi_{,\varphi}
\right]\bigg\vert_{\varphi=\varphi_*} = 0
\,,
\end{align}
which follows from Eqs. \eqref{eqn:bkgeom1} and \eqref{eqn:bkgeom2} at stationarity.
Note that $\xi_{,\varphi} \propto {\rm sech}^2[\xi_1(\varphi-\varphi_c)]$.
It is natural to choose $\varphi_c$ to be the fixed point, i.e., $\varphi_* = \varphi_c$. 
The condition $\varphi_* = \varphi_c$ is realized when
\begin{align}\label{eqn:xi1fp}
\xi_1 = \frac{6M_{\rm P}^4\sin(\varphi_c/f)}{f\Lambda^4\xi_0[1+\cos(\varphi_c/f)]^2}\,.
\end{align}
The second derivative of the Gauss-Bonnet coupling function $\xi$ with respect to the field $\varphi$ at the nontrivial fixed point $\varphi_*$ when $\xi_1$ takes Eq. \eqref{eqn:xi1fp} is given by $\xi_{,\varphi\varphi}(\varphi=\varphi_*) = 0$. Thus, the nontrivial fixed point becomes a saddle point, as pointed out in Ref. \cite{Kawai:2021bye}. Near the nontrivial fixed point, Eq. \eqref{eqn:bkgeom2} is approximated as $\ddot{\varphi}+3H\dot{\varphi} \approx 0$, indicating an ultra-slow-roll regime. 
An inflaton trajectory that passes near the nontrivial fixed point will enter an ultra-slow-roll regime and experiences an enhancement in the curvature power spectrum \cite{Kawai:2021bye}.

For a given inflationary background, the curvature perturbation $\zeta$ follows the equation
\begin{align}\label{eqn:perteomS}
v_{k}^{\prime\prime}
+\left(
C_{\zeta}^{2}k^{2}
-\frac{A_{\zeta}^{\prime\prime}}{A_{\zeta}}
\right)v_{k} = 0
\,,
\end{align}
where ${}^\prime$ denotes the conformal time $\tau$ derivative, and the quantity $v_{k}$ is $v_{k} \equiv M_{\rm P} A_{\zeta}\zeta_{k}$ with $\zeta_{k}$ being the Fourier mode of the curvature perturbation $\zeta$. The quantities $A_\zeta$ and $C_\zeta$ are given by
\begin{align}
A_\zeta^2 &=
a^2\left(
\frac{1-\sigma_1/2}{1-3\sigma_1/4}
\right)^2
\\
&\quad\times
\left(
2\epsilon_1 - \frac{1}{2}\sigma_1
+\frac{1}{2}\sigma_1\sigma_2
-\frac{1}{2}\sigma_1\epsilon_1
+\frac{3}{4}\frac{\sigma_1^2}{2-\sigma_1}
\right)
\,,\nonumber\\
C_\zeta^2 &=
1-\frac{a^{2}}{A_{\zeta}^{2}}\left(
\frac{\sigma_1}{2-3\sigma_1/2}
\right)^{2}
\\
&\quad\times
\left(
2\epsilon_1 + \frac{1}{4}\sigma_1 - \frac{1}{4}\sigma_1\sigma_2
-\frac{5}{4}\sigma_1\epsilon_1
\right)
\,.\nonumber
\end{align}
For derivations, see the Appendix of Ref. \cite{Kawai:2021bye}. 
Here, $\epsilon_i$ and $\sigma_i$ are defined as
\begin{align}\label{eqn:SRparams}
\epsilon_1 &\equiv -\frac{\dot{H}}{H^2}
\,,\;
\epsilon_{i>1} \equiv \frac{\dot{\epsilon}_{i-1}}{H\epsilon_{i-1}}
\,,\;
\sigma_1 \equiv \frac{H\dot{\xi}}{M_{\rm P}^2}
\,,\;
\sigma_{i>1} \equiv \frac{\dot{\sigma}_{i-1}}{H\sigma_{i-1}}
\,.
\end{align}

For the tensor mode, the perturbation equation is given by \cite{Kawai:2021bye}
\begin{align}\label{eqn:perteomT}
u^{\prime \prime}_k + \left(
C_t^2 k^2 - \frac{A_t^{\prime\prime}}{A_t}
\right)u_k = 0\,,
\end{align}
where $u_k$ is the Fourier transform of $u^{\pm}$ (the superscript $\pm$ indicates the two polarization modes, which are omitted hereafter) which is defined via
\begin{align}
h_{ij} = \frac{\sqrt{2}}{A_t M_{\rm P}}\sum_\pm u^\pm e_{ij}^\pm\,,
\end{align}
with $e_{ij}^\pm$ being the polarization tensor, and
\begin{align}
A_t^2 &=
a^2\left( 1-\frac{\sigma_1}{2} \right) \,,\\
C_t^2 &=
1 + \frac{a^2\sigma_1}{2A_t^2}\left(1-\sigma_2-\epsilon_1\right)
\,.
\end{align}

We numerically solve the background equations of motion, Eqs. \eqref{eqn:bkgeom1} and \eqref{eqn:bkgeom2}, and the perturbation equations Eqs. \eqref{eqn:perteomS} and \eqref{eqn:perteomT}. For the initial conditions for the perturbations, we adopt the standard Wentzel–Kramers–Brillouin (WKB) solutions on the Bunch-Davies vacuum,
\begin{align}
\lim_{\tau\rightarrow -\infty}v_{k}(\tau)
&= \frac{1}{\sqrt{2C_{\zeta}k}}e^{-iC_{\zeta} k\tau}
\,,\label{eqn:BDiniS}\\
\lim_{\tau\rightarrow -\infty}u_{k}(\tau)
&= \frac{1}{\sqrt{2C_{t}k}}e^{-iC_{t} k\tau}
\,.\label{eqn:BDiniT}
\end{align}
The power spectra for the curvature and tensor perturbations are then obtained as
\begin{align}
\mathcal{P}_\zeta &= \frac{k^3}{2\pi^2}|\zeta|^2
=\frac{k^3}{2\pi^2}\frac{|v_k|^2}{M_{\rm P}^2 A_\zeta^2}
\,,\label{eqn:PSs}\\
\mathcal{P}_t &= 2\times\frac{k^3}{2\pi^2}|h_k|^2 
= \frac{2k^3}{\pi^2}\frac{|u_k|^2}{M_{\rm P}^2 A_t^2}
\,,\label{eqn:PSt}
\end{align}
evaluated in the superhorizon limit.
The spectral index $n_s$ and the tensor-to-scalar ratio $r$ are then given by
\begin{align}\label{eqn:nsr}
n_s \equiv 1+\frac{d\ln\mathcal{P}_\zeta}{d\ln k}
\,,\quad
r \equiv \frac{\mathcal{P}_t}{\mathcal{P}_\zeta}\,,
\end{align}
evaluated at the pivot scale $k_* = 0.05 \; {\rm Mpc^{-1}}$.

\begin{figure}[h]
\includegraphics[width=8.5cm]{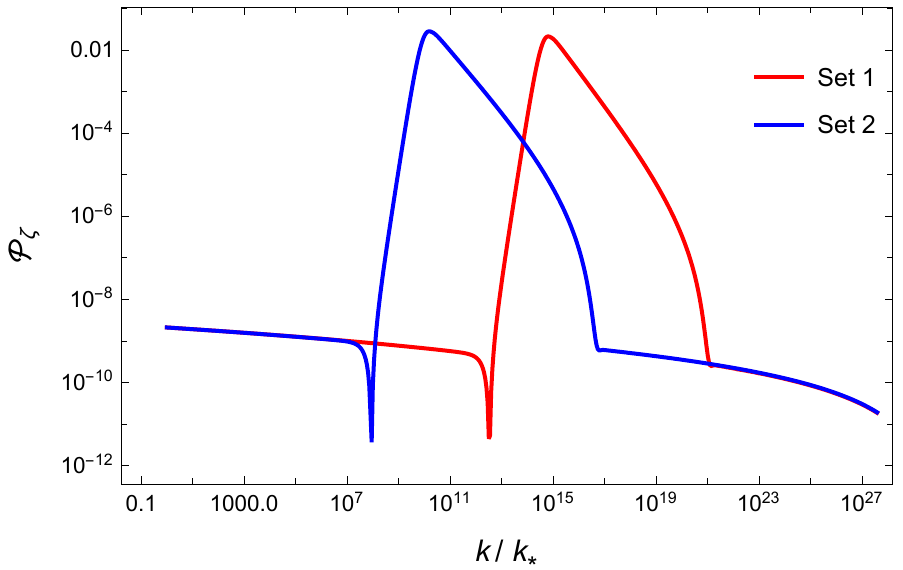}
\centering
\caption{The curvature power spectrum \eqref{eqn:PSs} is shown for our two benchmark parameter sets. The enhancement is observed as the inflaton enters the ultra-slow-roll regime near the nontrivial fixed point. Here, $k_* = 0.05\;{\rm Mpc}^{-1}$.}
\label{fig:cpsplot}
\end{figure}

In this work, we take the two benchmark parameter sets given in Table \ref{tab:BMsets}.
\begin{table}
\centering
\renewcommand{\arraystretch}{1.5}
\begin{tabular}{| c | c c c c c |} 
\hline
& $\Lambda \; (M_{\rm P})$ & $f \; (M_{\rm P})$ & $\varphi_c \; (M_{\rm P}) $ & $\xi_0 \; (10^7)$ & $\xi_1 \; (M_{\rm P}^{-1})$  \\ 
\hline\hline
Set 1 & 0.0065 & 7  & 13.0 & 6.044 & 15.0  \\ 
\hline
Set 2 & 0.0065 & 7 & 11.3 & 2.795 & 18.5  \\ 
\hline
\end{tabular}
\renewcommand{\arraystretch}{1}
\caption{Two sets of benchmark parameters.}
\label{tab:BMsets}
\end{table}
The value of $\Lambda$ is fixed from $A_s \approx 2.1\times10^{-9}$, where $A_s$ is the curvature power spectrum amplitude at the pivot scale $k_* = 0.05 \; {\rm Mpc^{-1}}$ \cite{Planck:2018jri}. For a given $f$, we are then left with three free parameters, $\xi_0$, $\xi_1$, and $\varphi_c$. The parameter $\varphi_c$ controls the peak position of the curvature perturbation, which is related to the size of the primordial blackholes and the peak frequency of the induced gravitational waves as shown in the next sections, while the parameter $\xi_1$ controls the width of the peak. The parameter $\xi_0$, which is responsible for the magnitude of the peak of the curvature perturbation, is chosen to be close to the value determined by Eq.~\eqref{eqn:xi1fp}.
Our numerical procedure is as follows. For a given parameter set, we solve the background equations of motion, \eqref{eqn:bkgeom1} and \eqref{eqn:bkgeom2}, with the assumption that the inflaton starts to roll down the potential \eqref{eqn:pot} near the origin; thus, inflation takes place in the range $0<\varphi<f\pi$, during which the inflaton is temporarily trapped near the $\varphi=\varphi_c$ point, and after that, it oscillates about and settles at $\varphi=f\pi$, where $V=0$. The end of inflation is found by using the condition $\epsilon_1 = 1$ where $\epsilon_1$ is the first Hubble slow-roll parameter defined as Eq. \eqref{eqn:SRparams}. Choosing the number of $e$-folds of 70 for the pivot scale, we find the horizon-crossing time of the mode. We then solve the perturbation equations, \eqref{eqn:perteomS} and \eqref{eqn:perteomT}, for a given wave number $k$. The initial conditions, \eqref{eqn:BDiniS} and \eqref{eqn:BDiniT}, are used when the mode is deep inside the horizon $k\gg aH$; concretely we require $k/(aH)\sim 10^{3}$. We evolve the perturbation equations until the mode exits the horizon and resides far outside the horizon $k\ll aH$; specifically, we require $k/(aH)\sim 10^{-3}$. We then use Eqs. \eqref{eqn:PSs} and \eqref{eqn:PSt} to compute the power spectra. For set 1 and set 2, the curvature power spectrum \eqref{eqn:PSs} is shown in Fig. \ref{fig:cpsplot}. As pointed out in Ref. \cite{Kawai:2021bye}, due to the ultra-slow-roll regime, we see a large enhancement in the curvature power spectrum.
Finally, we compute the spectral index $n_s$ and the tensor-to-scalar ratio $r$ by using Eq. \eqref{eqn:nsr}. They are given by $(n_s, r) = (0.96,0.079)$ for set 1 and $(0.96,0.080)$ for set 2.
At the time when the fluctuations seeding the primordial blackholes exit the horizon, the energy scale is $\rho^{1/4} \simeq 1.46 (1.58) \times10^{16}$ GeV for the benchmark parameter set 1 (2).

A large enhancement in the curvature power spectrum indicates the possibility of gravitational collapse when the corresponding mode reenters the horizon, which leads to the formation of primordial blackholes. Furthermore, such a large curvature perturbation may act as a source for the tensor perturbation at the nonlinear order as the scalar mode and the tensor mode are coupled. We discuss the production of primordial blackholes in the next section and the scalar-induced second-order tensor perturbation in Sec. \ref{sec:SIGW}.

\section{Production of primordial blackholes}
\label{sec:PBH}
Primordial blackholes may be produced due to the gravitational collapse when very large density fluctuations reenter the horizon. At its formation time $t_{\rm f}$, the mass of the produced primordial blackhole is given by
\begin{align}
M = \gamma M_{H,{\rm f}} = 4\pi\gamma\frac{M_{\rm P}^2}{H_{\rm f}}\,,
\end{align}
where $\gamma$ is the fraction factor that accounts for how much of the horizon mass turns into the primordial blackhole, and $M_H=4\pi M_{\rm P}^2/H$ is the horizon mass. The subscript ${\rm f}$ in $M_{H,{\rm f}}$ and $H_{\rm f}$ indicates a quantity at the formation time. We use $\gamma = 0.2$ as suggested by simple analytical estimations \cite{Carr:1975qj}.
From the entropy conservation, we find that the Hubble rate at the formation time, $H_{\rm f}$, is related to the Hubble rate at present, $H_0$, as
\begin{align}
H_{\rm f} = H_0 \frac{\Omega_{\rm rad,0}^{1/2}}{a_{\rm f}^2}\left(
\frac{g_{*,0}}{g_{*,{\rm f}}}
\right)^{\frac{1}{6}}\,,
\end{align}
where we have assumed that the formation occurs during the radiation-dominated era. Here, $a_{\rm f}$ is the scale factor at the formation time (the scale factor at present $a_0$ is set to be unity), $\Omega_{\rm rad} \equiv \rho_{\rm rad}/\rho_{\rm crit}$ the radiation energy density parameter with $\rho_{\rm crit}$ being the critical energy density of the Universe, and $g_*$ the effective relativistic degrees of freedom. Therefore, we find
\begin{align}
M = \gamma M_{H,0} \Omega_{\rm rad,0}^{1/2}\left(
\frac{g_{*,0}}{g_{*,{\rm f}}}
\right)^{\frac{1}{6}}\left(
\frac{k_0}{k_{\rm f}}
\right)^2\,.
\end{align}

The primordial blackholes behave as matter. 
Thus, they redshift as $\rho_{\rm PBH} \propto a^{-3}$. 
Then, we obtain $\rho_{\rm PBH,0} = \rho_{\rm PBH,f}(a_{\rm f}/a_0)^3 \approx \gamma\beta\rho_{\rm rad,f}(a_{\rm f}/a_0)^3$, where $\beta$ is the probability that the density fluctuation $\delta$ exceeds a threshold value $\delta_c$ for a given curvature power spectrum $\mathcal{P}_\zeta$, which is, assuming that the density fluctuation follows a Gaussian distribution,\footnote{
For the effects of non-Gaussianities on the primordial blackhole formation and the scalar-induced second-order gravitational waves, see, e.g., Refs. \cite{Bullock:1996at,Ivanov:1997ia,Lyth:2012yp,Byrnes:2012yx,Bugaev:2013vba,Young:2013oia,Nakama:2016gzw,Garcia-Bellido:2017aan,Franciolini:2018vbk,Cai:2018dig,Atal:2018neu,Unal:2018yaa,Passaglia:2018ixg,Atal:2019cdz,Panagopoulos:2019ail,Yoo:2019pma,Kehagias:2019eil,Ezquiaga:2019ftu,Yuan:2020iwf,Ragavendra:2020sop,Adshead:2021hnm,Atal:2021jyo}.
} given by
\begin{align}
\beta(M) = \int_{\delta_c} d\delta \frac{1}{\sqrt{2\pi \sigma^2}}\exp\left(
-\frac{\delta^2}{2\sigma^2}
\right)\,,
\end{align}
with the variance \cite{Josan:2009qn,Young:2014ana}
\begin{align}
\sigma^2 = \frac{16}{81}\int_0^\infty \frac{dq}{q} \left(\frac{q}{k}\right)^4 W^2\left(\frac{q}{k}\right)\mathcal{P}_\zeta(q)
\,.
\end{align}
We take the Gaussian window function $W(q/k) = \exp(-(q/k)^2/2)$ and use $\delta_c = 1/3$ \cite{Carr:1975qj}.

The total abundance of the primordial blackholes is given by $\Omega_{\rm PBH,tot} = \int d\ln M\; \Omega_{\rm PBH}$, where $\Omega_{\rm PBH}$ is conventionally expressed in terms of the quantity $f_{\rm PBH}$ given by
\begin{align}
f_{\rm PBH} &\equiv 
\frac{\Omega_{\rm PBH,0}}{\Omega_{\rm CDM,0}} =
\gamma^{\frac{3}{2}}\beta(M)\frac{\Omega_{\rm rad,0}^{3/4}}{\Omega_{\rm CDM,0}}\left(
\frac{g_{*,0}}{g_{*,{\rm f}}}
\right)^{\frac{1}{4}}\left(
\frac{M_{H,0}}{M}
\right)^{\frac{1}{2}}
\nonumber\\
&\approx
\Big(
\frac{\beta(M)}{3.27\times 10^{-8}}
\Big)\Big(
\frac{\gamma}{0.2}
\Big)^{\frac{3}{2}}\Big(
\frac{106.75}{g_{*,{\rm f}}}
\Big)^{\frac{1}{4}}
\nonumber\\
&\quad\times
\Big(
\frac{0.12}{\Omega_{\rm CDM,0}h^2}
\Big)\Big(
\frac{M}{M_\odot}
\Big)^{-\frac{1}{2}}
\,,
\end{align}
where $\Omega_{\rm CDM,0}$ is the current density parameter of the cold dark matter, $M_\odot$ the solar mass, and $h$ the rescaled present-day Hubble rate defined by $H_0=100\, h \;{\rm km/s/Mpc}$.
Here, we used the relation between the mass of the primordial blackhole $M$ and a scale $k$,
\begin{align}
M(k) \approx 4.64\times 10^{15}\, \gamma\, M_\odot\, \Big(
\frac{g_*}{106.75}
\Big)^{-1/6}\Big(
\frac{k}{k_*}
\Big)^{-2}\,,
\end{align}
where $k_* = 0.05\,{\rm Mpc}^{-1}$ is the pivot scale.

\begin{figure}[h]
\includegraphics[width=8.5cm]{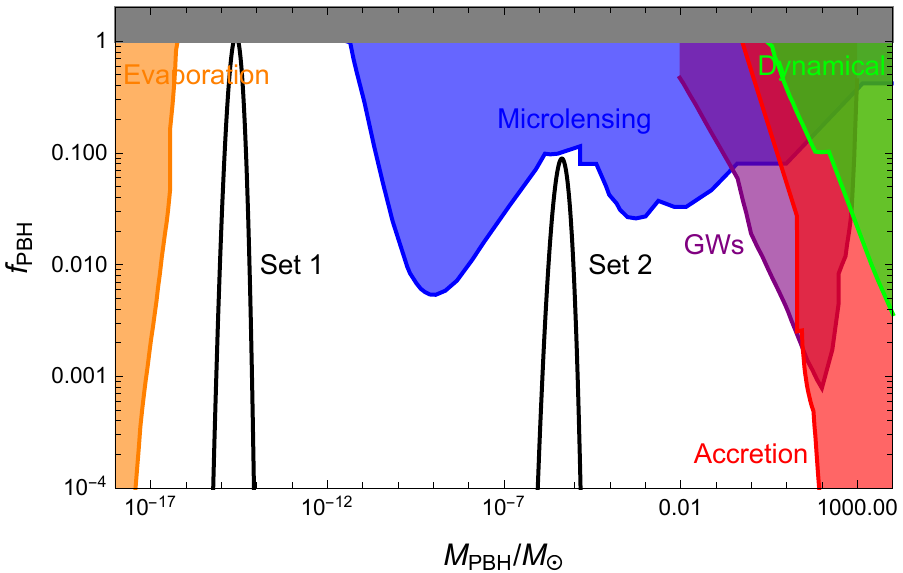}
\centering
\caption{The abundance of primordial blackholes is shown for our two benchmark parameter sets. Here, $f_{\rm PBH}\equiv \Omega_{\rm PBH,0}/\Omega_{\rm CDM,0}$. For the case of set 1, primordial blackholes constitute all of the dark matter abundance today. For the case of set 2, only a part of the present dark matter abundance is accounted for by the primordial blackholes. The data for the constraints are obtained from Refs. \cite{Green:2020jor,Kavanagh:2020aaa}. A recent analysis indicates more severe bounds from the gravitational waves \cite{Hutsi:2020sol}.}
\label{fig:pbhplot}
\end{figure}

In Fig. \ref{fig:pbhplot}, we present $f_{\rm PBH}$ for the two sets of parameter values, together with the current constraints \cite{Green:2020jor,Kavanagh:2020aaa}. For set 1, we obtain $f_{\rm PBH}^{\rm tot} \approx 1$, where $f_{\rm PBH}^{\rm tot} \equiv \int d\ln M f_{\rm PBH}$. Thus, the primordial blackholes may constitute all of the dark matter abundance today.
For set 2, we have $f_{\rm PBH}^{\rm tot} \approx 0.087$.

\section{Detectability of induced gravitational waves}
\label{sec:SIGW}
The enhanced curvature perturbation may become a source for the tensor perturbation at the nonlinear order \cite{Matarrese:1997ay,Mollerach:2003nq,Ananda:2006af,Baumann:2007zm}; see also Refs. \cite{Noh:2004bc,Hwang:2007ni,Gong:2019mui,Kohri:2018awv,Domenech:2019quo}. It is due to the fact that the scalar mode and the tensor mode couple to each other at the second order in perturbations.
To study the tensor perturbation induced by the second-order scalar perturbation, we work in the conformal Newtonian gauge,\footnote{
For the issue of gauge dependence, see, e.g., Refs. \cite{Hwang:2017oxa,DeLuca:2019ufz,Inomata:2019yww,Yuan:2019fwv,Domenech:2020xin}.
} where the metric is given by
\begin{align}
ds^2 = -(1+2\Phi)dt^2 + a^2\left[
(1-2\Psi)\delta_{ij} + \frac{1}{2}h_{ij}
\right]dx^i dx^j\,,
\end{align}
with $\Phi$ and $\Psi$ being scalar perturbations, and $h_{ij}$ the transverse-traceless tensor perturbation to the second order. 
We neglect the first-order contribution to the tensor perturbation below. Furthermore, we neglect the vector perturbation and the anisotropic stress tensor, and thus, $\Phi = \Psi$. 
The energy density of the gravitational waves in the subhorizon region is given by \cite{Maggiore:1999vm,Maggiore:2007ulw}
\begin{align}\label{eqn:rhoGW}
\rho_{\rm GW} = \frac{M_{\rm P}^2}{16a^2}\langle
\overline{\partial_k h_{ij} \partial^k h^{ij}}
\rangle\,,
\end{align}
where the overline indicates the average over oscillations, and the angle brackets denote the expectation value. In the Fourier space, the tensor mode $h_{ij}$ can be decomposed into two polarization modes,
\begin{align}\label{eqn:FTh}
h_{ij}(t,{\bf x}) = \int \frac{d^3k}{(2\pi)^{3/2}}\left(
h_{\bf k}^+(t)e_{ij}^+({\bf k}) + h_{\bf k}^\times(t)e_{ij}^\times({\bf k})
\right)e^{i{\bf k}\cdot{\bf x}}\,,
\end{align}
where the polarization tensors are given by
\begin{align}
e_{ij}^+({\bf k}) &= \frac{1}{\sqrt{2}}\left(
e_i({\bf k})e_j({\bf k}) - \bar{e}_i({\bf k})\bar{e}_j({\bf k})
\right) \,,\\
e_{ij}^\times({\bf k}) &= \frac{1}{\sqrt{2}}\left(
e_i({\bf k})\bar{e}_j({\bf k}) + \bar{e}_i({\bf k})e_j({\bf k})
\right)\,,
\end{align}
with $e_i({\bf k})$ and $\bar{e}_i({\bf k})$ being two orthogonal unit vectors that are perpendicular to ${\bf k}$.
Substituting Eq. \eqref{eqn:FTh} into Eq. \eqref{eqn:rhoGW}, we obtain
\begin{align}
\rho_{GW}(t) = \int d\ln k \, \frac{M_{\rm P}^2}{8}\left(\frac{k}{a}\right)^2
\overline{\mathcal{P}_h(t,k)}\,,
\end{align}
where $\mathcal{P}_h(t,k) \equiv \mathcal{P}^{+,\times}_h(t,k)$ is the tensor power spectrum, defined by
\begin{align}
\langle h_{\bf k}^+(t)h_{\bf q}^+(t) \rangle &=
\delta^3({\bf k} + {\bf q})\frac{2\pi^2}{k^3}\mathcal{P}^+_h(t,k)\,,\\
\langle h_{\bf k}^\times(t)h_{\bf q}^\times(t) \rangle &=
\delta^3({\bf k} + {\bf q})\frac{2\pi^2}{k^3}\mathcal{P}^\times_h(t,k)\,.
\end{align}
Note that $\mathcal{P}^+_h(t,k) = \mathcal{P}^\times_h(t,k)$ in the absence of $CP$ violation. In the following, we omit the polarization index.
It is conventional to define the energy density parameter of the gravitational waves as
\begin{align}
\Omega_{\rm GW}(t,k) \equiv \frac{\rho_{\rm GW}(t,k)}{\rho_{\rm crit}}
= \frac{1}{24}\left(\frac{k}{aH}\right)^2\overline{\mathcal{P}_h(t,k)}\,,
\end{align}
where $\rho_{\rm crit}$ is the critical energy density of the Universe.

The tensor power spectrum is obtained by solving the equation of motion for the tensor perturbation which is given by the transverse-traceless component of the second-order Einstein equation. The equation of motion is given, in the Fourier space, by \cite{Baumann:2007zm,Kohri:2018awv,Inomata:2016rbd}
\begin{align}
h_{\bf k}^{\prime\prime} + 2\mathcal{H}h_{\bf k}^\prime + k^2h_{\bf k} = 4S_{\bf k}\,,
\end{align}
where $\mathcal{H}\equiv a^\prime/a = aH$ and $S_{\bf k}$ is the source term,
\begin{align}
S_{\bf k} &= \int \frac{d^3q}{(2\pi)^{3/2}} e_{ij}({\bf k})q_iq_j\bigg[
2\Phi_{\bf q}\Phi_{{\bf k}-{\bf q}}
\nonumber\\
&\quad
+\frac{4}{3(1+w)}(\mathcal{H}^{-1}\Phi_{\bf q}^\prime + \Phi_{\bf q})
(\mathcal{H}^{-1}\Phi_{{\bf k}-{\bf q}}^\prime + \Phi_{{\bf k}-{\bf q}})
\bigg]\,.
\end{align}
Here, $w$ is the equation of state parameter. 
The solution can be obtained by means of Green's function.
We assume that the scalar-induced second-order gravitational waves are produced during the radiation-dominated era. The energy density parameter at the production time, $\Omega_{\rm GW,f}(k) = \Omega_{\rm GW}(t_{\rm f},k)$, is given by \cite{Kohri:2018awv}
\begin{align}\label{eqn:OmegaGWf}
&\Omega_{\rm GW,f}(k) =
\frac{1}{12}\int_0^\infty dv
\int_{|1-v|}^{1+v} du
\left(
\frac{4v^2 - (1+v^2-u^2)^2}{4uv}
\right)^2
\nonumber\\
&\quad\times
\mathcal{P}_\zeta(kv)\mathcal{P}_\zeta(ku)
\left(
\frac{3(u^2+v^2-3)}{4u^3v^3}
\right)^2
\nonumber\\
&\quad\times
\bigg[
\left(
-4uv + (u^2+v^2-3)\log\bigg\vert
\frac{3-(u+v)^2}{3-(u-v)^2}
\bigg\vert
\right)^2 
\nonumber\\
&\hspace{2cm}
+ \pi^2(u^2+v^2-3)^2\theta(v+u-\sqrt{3})
\bigg]\,.
\end{align}
The density parameter today is then given by \cite{Kohri:2018awv} (see also Ref. \cite{Ando:2018qdb})
\begin{align}\label{eqn:OmegaGW0}
\Omega_{\rm GW} = \Omega_{\rm rad,0}\Omega_{\rm GW,f}\,,
\end{align}
where $\Omega_{\rm rad,0} \approx 0.9\times 10^{-4}$ is the current energy density parameter of radiation.

\begin{figure}[h]
\includegraphics[width=8.5cm]{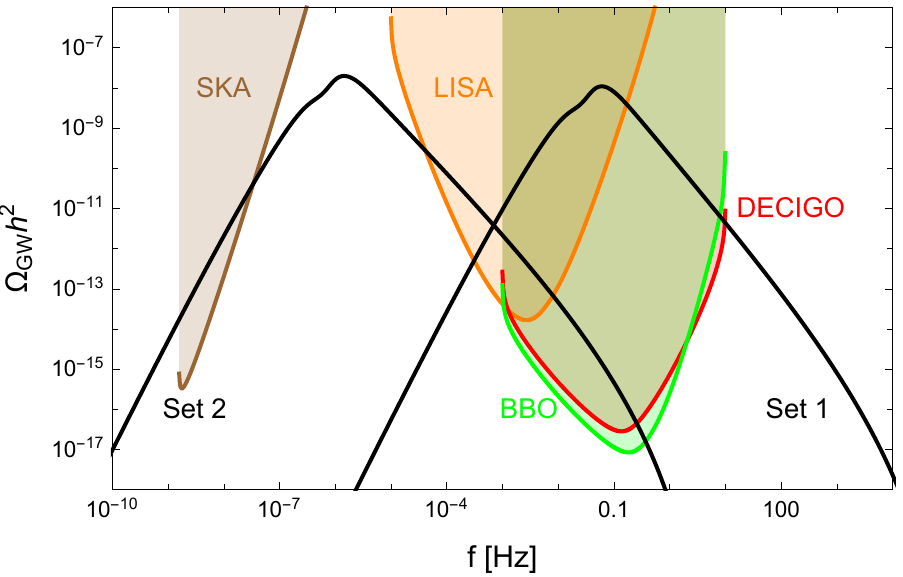}
\centering
\caption{The density parameter of the scalar-induced second-order gravitational waves is shown for our two benchmark sets. The gravitational wave signal of set 1 is well within the reach of the sensitivity bound of future experiments such as LISA, DECIGO, and BBO. In the case of set 2, the signal crosses the sensitivity bound of SKA as well. The data for the sensitivity curves are obtained from Refs. \cite{Schmitz:2020syl,Schmitz:2020aaa}.}
\label{fig:sigwplot}
\end{figure}

Utilizing Eqs. \eqref{eqn:OmegaGWf} and \eqref{eqn:OmegaGW0}, together with the curvature power spectrum in Fig. \ref{fig:cpsplot}, we present the gravitational waves density parameter in Fig. \ref{fig:sigwplot} for our two sets of parameter values.
The gravitational wave signal of set 1 is within the reach of future experiments such as LISA \cite{LISA:2017pwj,Baker:2019nia}, DECIGO \cite{Seto:2001qf,Kawamura:2006up}, and BBO \cite{Crowder:2005nr,Corbin:2005ny,Harry:2006fi}. For set 2, the signal crosses the sensitivity bound of Square Kilometer Array (SKA) \cite{Carilli:2004nx,Janssen:2014dka,Weltman:2018zrl} in addition to that of LISA, DECIGO, and BBO.

\section{Final remarks}
\label{sec:conc}
We studied the production of primordial blackholes and the scalar-induced second-order gravitational wave signals in a model where a scalar (inflaton) is coupled to the Gauss-Bonnet term.
The presence of the Gauss-Bonnet coupling function indicates the existence of a nontrivial de Sitter-like fixed point. Near the nontrivial fixed point the inflaton enters an ultra-slow-roll regime. During the ultra-slow-roll inflation period the curvature power spectrum gets enhanced.
We considered two benchmark parameter sets and showed that a large enhancement occurs in the curvature power spectrum by numerically solving the equations of motion.

A mode with large enhancement of the curvature perturbation may experience gravitational collapse when reentering the horizon, thereby producing primordial blackholes. For our two benchmark sets, we computed the present abundance of primordial blackholes. One set accounts for the totality of the dark matter relic density today, while in the other case primordial blackholes constitute only a portion of the present dark matter relic abundance.

A large curvature perturbation that leads to the production of primordial blackholes inevitably source the scalar-induced second-order gravitational waves. The present density parameter of the gravitational waves is obtained by utilizing the approximated analytical expression together with our numerical results of the curvature power spectrum. Both of our two benchmark sets are found to be within the sensitivity bounds of future gravitational wave experiments such as LISA, DECIGO, BBO, and SKA.

While we focused on the scalar potential of the natural inflation model and assumed a smeared step function for the Gauss-Bonnet coupling function in this work, some of the features that we have found are generic. When there is a balance between a scalar potential term and a Gauss-Bonnet coupling term, a nontrivial fixed point may exist. Near the nontrivial fixed point the ultra-slow-roll inflation generically occurs, during which period a large enhancement of the curvature perturbation is guaranteed. We thus expect that the production of primordial blackholes and the secondary gravitational wave signals are natural in higher curvature gravity theories.

\begin{acknowledgments}

We acknowledge helpful communications with Kazunori Kohri and useful discussions with Valerie Domcke and Kai Schmitz.
This work was supported in part by the National Research Foundation of Korea Grant-in-Aid for Scientific Research Grant No.
NRF-2018R1D1A1B07051127 (S.K.).
\end{acknowledgments}




\end{document}